\documentclass[twocolumn,,amssymb, nobibnotes, aps, prd, nofootinbib]{revtex4-2}

\usepackage{amsmath}
\usepackage{amsfonts}
\usepackage{amssymb}
\usepackage{booktabs}
\usepackage[T1]{fontenc}
\usepackage{graphicx}
\usepackage[colorlinks=true, linkcolor=blue, urlcolor=blue, citecolor=blue]{hyperref}
\usepackage{siunitx}
\usepackage{xcolor}

\newcommand{\abs}[1]{\left|#1\right|}
\newcommand{\bsh}{\texttt{BinarySkyHough} }
\newcommand{\coincth}{d^{\textrm{th}}}
\newcommand{\F}{\mathcal{F}}
\newcommand{\hal}{h_{\mathcal{A}, \lambda}}
\newcommand{\reachth}{d_{f_0}^{\textrm{th}}}
\newcommand{\sh}{\texttt{SkyHough} }
\newcommand{\Tsft}{T_{\text{SFT}}}

\graphicspath{{figures/}}

\begin{document}

\title{Time-frequency track distance for comparing continuous gravitational wave signals}

\author{Rodrigo Tenorio} 
\email{rodrigo.tenorio@ligo.org}
\affiliation{Departament de F\'isica, Institut d'Aplicacions Computacionals i de Codi Comunitari (IAC3), Universitat de les Illes Balears, 
and Institut d'Estudis Espacials de Catalunya (IEEC), Crta. Valldemossa km 7.5, E-07122 Palma, Spain}
\author{David Keitel}
\affiliation{Departament de F\'isica, Institut d'Aplicacions Computacionals i de Codi Comunitari (IAC3), Universitat de les Illes Balears, 
and Institut d'Estudis Espacials de Catalunya (IEEC), Crta. Valldemossa km 7.5, E-07122 Palma, Spain}
\author{Alicia M. Sintes}
\affiliation{Departament de F\'isica, Institut d'Aplicacions Computacionals i de Codi Comunitari (IAC3), Universitat de les Illes Balears, 
and Institut d'Estudis Espacials de Catalunya (IEEC), Crta. Valldemossa km 7.5, E-07122 Palma, Spain}

\date{\today}%

\begin{abstract}
    Searches for continuous gravitational waves from unknown sources attempt to detect long-lasting 
    gravitational radiation by identifying Doppler-modulated signatures in the data.
    Semicoherent methods allow for wide parameter space surveys, identifying interesting
    regions to be followed up using more sensitive (and computationally expensive) tools. Thus, it is required
    to properly understand the parameter space structure under study, as failing to do so could significantly
    affect the effectiveness of said strategies. We introduce a new measure for distances in parameter space
    suited for semicoherent continuous wave searches. This novel approach, based on comparing time-frequency 
    tracks, can be applied to any kind of quasi-monochromatic continuous wave signals and adapts itself to the 
    underlying structure of the parameter space under study.
    In a first application to the post-processing stage of an all-sky search for continuous waves
    from neutron stars in binary systems, we demonstrate a search sensitivity improvement by solely replacing
    previous ad hoc distance measures in the candidate clustering procedure by the new proposal.
\end{abstract}

\maketitle

\section{Introduction}

Continuous gravitational waves (CWs) are long-lasting, quasi-monochromatic, gravitational wave (GW)
signals emitted by sustained quadrupolar variations of a mass distribution. As opposed to
GWs from compact binary coalescences~\cite{Abbott:2020niy}, CWs
require long periods of data to be integrated in order to accumulate a significant signal-to-noise ratio (SNR),
as expected signals are very faint. Focusing on current ground-based detectors like Advanced LIGO
\cite{AdvancedLIGO} and Advanced Virgo \cite{AdvancedVirgo}, possible
sources include rapidly-spinning non-axisymmetric neutron stars, which could emit GW
radiation through a variety of mechanisms (see \cite{Sieniawska_2019} for a recent review), as well
as more exotic alternatives such as ultralight boson clouds around spinning black holes (see 
\cite{PhysRevD.102.063020} and references therein).

Several methods have been proposed to search for CWs, depending on the available information about the 
targeted sources. On one end, targeted searches aim at known pulsars, assuming a
tight phase locking between electromagnetic and gravitational radiation; on the other end, all-sky 
searches try to uncover unknown sources, imposing fewer constraints on possible signals.

Narrow parameter-space regions, as considered by targeted and other well-constrained searches, can 
be effectively analyzed using coherent methods which compare the full data stream to a bank of 
phase-evolution templates.
The number of templates to be considered in a coherent search scales with a strong power
of the observing time \cite{Prix:2006wm}. As a result, these methods can not be directly applied to wider 
parameter-space surveys, such as all-sky searches.

Semicoherent methods (e.g.\;\cite{PhysRevD.70.082001, Cutler:2005pn, 2014PhRvD..90d2002A, DAntonio:2018sff, BSH}),
on the other hand, divide the data stream into a discrete set of segments, each delivering a coherent statistic 
taking the full phase evolution into account. The final detection statistic is then computed incoherently
from these coherent statistics by following the frequency-evolution track associated to each
parameter-space candidate across segments.
Since phase and frequency evolution are related by time differentiation, a single frequency-evolution
template is related to an ensemble of phase-evolution templates, effectively reducing the number of required
templates to cover a parameter space region (see e.g. \cite{Pletsch:2010xb, Leaci:2015bka}).
Also, this renders semicoherent methods insensitive to discrete phase jumps between coherent
segments. Consequently, these types of searches are susceptible to deliver more spurious candidates
than coherent searches, which then require further investigation \cite{Dergachev:2019wqa}.

Typical semicoherent searches follow a hierarchical scheme, applying increasingly more sensitive 
(and more computationally expensive) stages as interesting parameter-space regions are narrowed down
\cite{Prix:2012yu}. Both the setup of such hierarchical searches and
the post processing of any outlier candidates depend on an understanding of the parameter-space structure.
For example, many searches \cite{PhysRevD.96.122004, O1AllSky, O1FullBandAllSky, PhysRevD.100.024004, 
PhysRevLett.124.191102, Steltner:2020hfd} use clustering algorithms to reduce the number of 
correlated candidates that proceed from one stage to the next. Most clustering algorithms depend on a
meaningful measure of parameter-space distance to represent said correlations within the
algorithm.  

Much of the understanding about CW parameter-space correlations has been built upon the $\F$-statistic,
a maximum likelihood estimator accounting for the SNR in a coherent CW search \cite{JKS1998, PhysRevD.72.063006}.
For a signal present in some data, the best-matching template will return a high $\F$-statistic value.
Neighboring templates, however, will also show enhanced (albeit smaller) $\F$-statistic values.
The relative loss due to mismatched templates was used to propose a family of local parameter-space
metrics \cite{Prix:2006wm}. These metrics can be used to set up optimal parameter-space grids, placing
templates in such a way that any signal within the considered parameter space is not mismatched by more than a
certain fixed amount \cite{Wette:2013wza, Wette:2014tca, Allen:2019vcl}. 

Optimal template-placement methods have been successfully generalized to semicoherent methods
\cite{Pletsch:2010xb, Wette:2015lfa, Wette:2018bhc}, improving the overall effectiveness of hierarchical schemes.
However, there is currently still a lack of literature on the definition of distance measures for the
post-processing of candidates from semicoherent search stages.

In this paper we propose a new parameter-space distance suited for the post-processing of semicoherent
CW searches by comparing the average mismatch along the frequency evolution tracks associated to a pair
of parameter-space points. This novel approach can be applied to any kind of quasi-monochromatic CW 
signal as long as a model of its frequency evolution is available. 
We provide a theoretical justification for this proposal 
by relating it to the \mbox{$\F$-statistic}. As a demonstration of its usefulness, we apply it to a clustering 
algorithm, comparing its effectiveness with respect to the results obtained in a previous semicoherent search.

In Sec.~\ref{sec:current} we review how parameter-space structure is taken into account in current
post-processing approaches. We introduce and discuss the properties of our proposal in Sec.~\ref{sec:newdistance}.
In Sec.~\ref{sec:clustering} we discuss a clustering algorithm as a first application of the new distance
and give a numerical assessment of its effectiveness in Sec.~\ref{sec:comparison} by improving the sensitivity
of an all-sky search using open data from the second observing run of the Advanced LIGO detectors.
We summarize our proposal and conclusions in Sec.~\ref{sec:conclusion}.

\section{Parameter-space distances and clustering algorithms}
\label{sec:current}

The main application of parameter-space distances in the context of semicoherent searches is to reduce
the number of outliers by grouping them into clusters,
thus reducing the computational cost and human effort of additional follow-up stages.
This can also shift the focus from individual candidates to correlated parameter-space regions,
and help associate these to specific origins such as instrumental noise or astrophysical signals.

To understand the potential benefits of our new proposed distance measure,
here we briefly review the ways in which parameter-space structure is taken into
account by clustering algorithms currently in practical use for CW all-sky searches.
Those can be divided into two families according to the way in which they take
parameter-space information into account, either based on fixed grids or on machine learning.

\subsection{Grid-based clustering methods}
\label{sec:gridclustering}

The first family of methods relies on the definition of a distance in terms of a parameter-space grid.
This approach measures \emph{how many grid steps away} two given parameter-space points are from each other,
making it well suited to estimate the computing costs of further follow-up stages.

Examples of this approach are the all-sky searches for CWs from isolated neutron stars in LIGO O1 and O2 data
\cite{O1AllSky, O1FullBandAllSky, PhysRevD.100.024004}. The signals covered by these searches can be
described by four parameters, namely initial frequency, spin-down and sky position:
$\lambda = \{f_0, f_1, \vec{n}\}$. 

The Euclidean parameter-space distance is defined as
\begin{equation}
    d(\lambda, \lambda_{*}) = \sqrt{ 
    \left( \frac{\Delta f_{0}}{\delta f_{0}}\right)^2
    + \left( \frac{\Delta f_{1}}{\delta f_{1}}\right)^2
    + \left( \frac{\Delta \theta}{\delta \theta}\right)^2
    }\;,
    \label{eq:euclidean}
\end{equation}
where $\Delta$ quantities represent one-dimensional differences between parameter-space points and $\delta$
quantities represent grid resolutions, which are constructed according to the criteria of each search pipeline.
While the frequency and spin-down contributions are treated rather consistently, different pipelines have opted
for different prescriptions for the sky contribution $\theta$. A common characteristic amongst O1 searches was 
the choice of the \emph{great circle distance} to account for the sky contribution in Eq.~\eqref{eq:euclidean}. 

As discussed in \cite{PhysRevD.70.082001}, the structure of the sky subspace is highly anisotropic, as CW signals
located close to the ecliptic plane tend to be more delocalized than those coming from the ecliptic poles. This
is due to the Doppler modulation induced by the Earth's motion, the dominant component of which is given by the
yearly motion around the Solar System Barycenter (SSB).
Thus, an alternative ansatz was proposed in \cite{AdaptativeClustering} by
projecting sky positions onto the ecliptic plane
\begin{equation}
    \Delta \theta^{2} = \Delta x^2 + \Delta y^2\;,
    \label{eq:ecliptic}
\end{equation}
where $\Delta x, \Delta y$ represent the difference in Cartesian ecliptic coordinates. A convenient anisotropic 
sky grid is implicitly set up after neglecting contributions from the third Cartesian component $\Delta z$:
The sky grid around the poles remains untouched, while the region around the ecliptic plane becomes \emph{wider} 
along the polar direction, in qualitative agreement with the described sky structure.
This idea was applied to the \sh pipeline during the LIGO O2 all-sky search
\cite{PhysRevD.100.024004}. However the great circle distance was still used to estimate local sky distances for
the setup of follow-up stages~\cite{PhysRevD.100.024004, PhysRevLett.124.191102}.

Extensions to Eq.~\eqref{eq:euclidean} to account for the additional modulations experienced by signals from 
sources in circular binary orbits were proposed in \cite{PhysRevLett.124.191102}, adapting this type of methods 
to the search for CWs from neutron stars in binary systems.
The proposal considered the three parameters required to describe a circular binary orbit,
namely the projected semi-major axis $a_{p}$, orbital frequency $\Omega$, and time of passage through the 
ascending node $t_{\textrm{asc}}$, adding them in quadrature as an extra term to Eq.~\eqref{eq:euclidean}:
\begin{equation}
    d_{\textrm{Binary}}(\lambda, \lambda_{*}) = \sqrt{ 
    \left( \frac{\Delta a_p}{\delta a_p} \right)^2
    + \left( \frac{\Delta \Omega}{\delta \Omega} \right)^2
    + \left( \frac{\Delta t_{\textrm{asc}}}{\delta t_{\textrm{asc}}} \right)^2
    }\;.
    \label{eq:euclidean_binary}
\end{equation}

Current clustering applications to CW searches then typically group candidates located within a certain
distance $\coincth$ of each other, with $\coincth$ calibrated to maximize the detection efficiency for a
population of artificial signals injected into the data. For the searches referenced above, this was chosen
as $\coincth \in [3,\;4]$ parameter-space bins.

The effectiveness of these clustering methods has been demonstrated by application in the cited searches.
It is unclear, however, how to extend them towards a more general description of the parameter space.
So far, distance proposals have assumed parameters to be uncorrelated. But this is generally not true, with correlations
highly dependent on the parameter space under analysis \cite{Pletsch:2010xb, Leaci:2015bka, Wette:2015lfa}.
Also regarding the actual functional form in which each parameter-space dimension enters into the
equation, existing grid-based proposals turned out to be effective for ``simple'' cases, such as isolated sources
(e.g. Eq. \eqref{eq:ecliptic}). However, more complex types of signals, such as those modulated by
binary orbits, pose a challenge to this approach.

\subsection{Machine Learning-based clustering}\label{sec:mlclustering}

The second family of clustering methods uses machine learning (ML) techniques to classify parameter-space
regions and identify those potentially containing CW signals \cite{Morawski_2020, DeepLearningClustering, 
deepLearningClusteringII}.
The input is usually a graphical representation of the search outputs over the parameter space, e.g.
the detection statistic values obtained during the main stage of the search.
Algorithms then learn to classify these through the usual train-test-validate
ML workflow  (see e.g. \cite{Goodfellow-et-al-2016}).

This approach presents several advantages with respect to methods from Sec.~\ref{sec:gridclustering}, as
ML classifiers are able to learn parameter-space correlations through the provision of representative
realizations of parameter-space regions during the training stage of the algorithm. These training examples are
relatively easy to produce, making it computationally feasible to train ML classifiers using a rich set of correlation
examples. In this sense, the parameter-space structure is no longer described by an explicit distance function.
Still, one has to carefully consider the way in which search results are represented in order to properly train the
algorithm and hence to fully exploit the potential of ML techniques.

\section{A new parameter space distance}
\label{sec:newdistance}

We introduce a new parameter space distance with the intent of overcoming the limitations identified during the
previous section. We employ the usual description of CWs in terms of the instantaneous
frequency evolution of a signal, which is in turn described in terms of the physical parameters of the source.
By considering frequency evolution tracks rather than a particular choice of parameters, we are able to
describe a general distance that relates directly to the same information content exploited by semicoherent CW searches.

\subsection{Deriving a distance from the $\F$-statistic}\label{sec:Fstar}

CWs can be characterized by two sets of parameters, namely the phase 
parameters $\lambda$ and the amplitude parameters $\mathcal{A}$. The former contain information 
about the phase and frequency evolution of a signal, which is due to both physical mechanisms intrinsic to the source
(e.g. binary orbital motions) and relative motions of a GW detector with respect to the
SSB; the latter describe the amplitude of a signal, encoding information about
the GW polarization, as well as the antenna response of the detector.

The detector response to a CW can be expressed as a linear superposition of four filters,
obtained by projecting each GW polarization onto both detector response functions:
\begin{equation}
    \hal(t) = \sum_{\mu = 0}^{3} \mathcal{A}^{\mu} h_{\mu}(t;\lambda)
    \label{eq:JKS} 
\end{equation}
where each component $\mathcal{A}^{\mu}$ is a time-independent expression depending only on the amplitude 
parameters and $h_{\mu}$ depends on the phase parameters only~\cite{JKS1998,Whelan:2013xka}.

Let us now assume a time series $x(t)$ consisting of additive Gaussian noise and a CW signal 
$\hal(t)$. Then, following the Neyman-Pearson criterion \cite{JKS1998, doi:10.1098/rsta.1933.0009}, 
the optimal detection statistic would be given by the likelihood ratio
\begin{equation}
\label{eq:lnL} 
    \log{\Lambda\left(\mathcal{A}, \lambda\right)} \propto \langle x, \hal \rangle\;,
\end{equation} 
where $\langle \cdot, \cdot \rangle$ represents a noise-weighted scalar product comparing
data to CW templates (high values corresponding to good alignment between data stream and
the proposed template). Since amplitude parameters enter as linear coefficients in Eq.~\eqref{eq:JKS},
analytic maximization can be performed in Eq.~\eqref{eq:lnL},
obtaining the so-called $\F$-statistic~\cite{JKS1998, PhysRevD.72.063006}
\begin{equation} 
    \label{eq:F} 
    \F(\lambda) 
    \equiv \max_{\mathcal{A}}\log{\Lambda}\left(\mathcal{A}, \lambda\right)\;.
\end{equation}
However, as first pointed out in \cite{Prix_2005}, this statistic presents a certain degeneracy with respect
to phase parameters; in other words, the locus of phase parameters for which Eq.~\eqref{eq:F} attains (close to) a
maximum value is not simply a point, but a finite parameter-space region.

A study of this property was done in \cite{PhysRevD.78.102005} using a simplified version of Eq.~\eqref{eq:JKS} by
neglecting amplitude modulations and solely focusing on the phase parameter dependency 
\begin{equation}
    \label{eq:h_star} 
    \hal^{\star}(t) = A \sin\Phi_{\lambda}(t) + B \cos\Phi_{\lambda}(t)\;,
\end{equation}
where $\Phi_{\lambda}$ is the phase evolution associated to $\lambda$.
The validity of this model and its agreement with Eq.~\eqref{eq:F} was justified in \cite{JK1999};
in turn, it implies a corresponding $\F^\star$ statistic by maximizing with respect to the 
(now constant) amplitude parameters:
\begin{equation} 
    \label{eq:F_star} 
    \F^{\star}(\lambda) 
    \equiv \max_{A, B}\log{\Lambda^{\star}} \propto \abs{\mathcal{X}(\lambda)}^{2}\;,
\end{equation}
where the following notation for the simplified filter was introduced:
\begin{equation} 
    \label{eq:X}
    \mathcal{X}(\lambda) = \langle x, e^{-i \Phi_{\lambda}}\rangle\;.  
\end{equation}

Let us now assume a CW signal described by a set of parameters $\lambda_{*}$ is present in our data.
If we neglect rapidly oscillating terms in Eq.~\eqref{eq:X}, we can express the $\F^{\star}$
statistic in terms of the phase difference between the template parameters $\lambda$ and 
those of the signal, \mbox{$\Delta \Phi(t) = \Phi_{\lambda_{*}}(t) - \Phi_{\lambda}(t)$}, namely
\begin{equation}
    \label{eq:X_approx} 
    \mathcal{X}(\lambda) \propto \int_{0}^{T} \textrm{d}t \; e^{i \Delta \Phi(t)}\;.
\end{equation}
As discussed in \cite{PhysRevD.78.102005}, regions where $\F^{\star}$ attains a maximum value 
will approximately correspond to those for $\F$. It is clear from Eq.~\eqref{eq:F_star} that a maximum
value of $\F^{\star}$ corresponds to a maximum value of $\abs{\mathcal{X}}$, which is achieved 
whenever
\begin{equation}
    \label{eq:condition}
    \Delta \Phi(t) = 0 \quad \forall t\;.
\end{equation}
We can re-state this condition in terms of the instantaneous frequency of said pair of templates 
$f_{\lambda}(t)$
\begin{equation}
    \frac{\partial \Delta \Phi}{\partial t} = 0 \rightarrow 
    f_{\lambda_{*}}(t) - f_{\lambda}(t) = 0 \quad \forall t\;,
    \label{eq:freq_zero}
\end{equation}
which is to say the statistic will be maximal for those pairs of templates whose frequency evolution 
coincides over the observing time. We also refer to the frequency evolution of a candidate or template as its
\emph{time-frequency track}, or simply \emph{track}.

With Eq.~\eqref{eq:freq_zero} we start our new proposal: the information a search recovers is best described not
just by the parameter-space point corresponding to a candidate, but by its actual frequency evolution.
As a result, we can relate different parameter-space points by comparing their frequency evolution tracks using an
arbitrary functional distance
\begin{equation}
    d(\lambda_{*}, \lambda) \equiv \mathfrak{D}\left[f_{\lambda_{*}} - f_{\lambda}\right]\;.
    \label{eq:primal_tbd}
\end{equation}
This means the distance amongst parameter-space candidates is measured in a consistent way with
semicoherent searches, where the statistical significance is defined as an integral along
a time-frequency track. In the same sense a frequency-evolution mismatch would produce a loss in
significance, it accounts for the amount of distance from one candidate to another.

For the remainder of this work we choose to focus on the $L_{1}$ distance in order to reduce the amount 
of implicit weights involved in our analysis, noting that any other proposal from the $L_{p}$ family
should yield similar results.

Before concluding this exposition, it is worth noticing there were no assumptions on the actual form of
$f_{\lambda}$ except for the fact it had to be a CW signal (i.e. long-lasting and quasi-monochromatic).
This means that Eq.~\eqref{eq:primal_tbd} could
be used for any kind of quasi-monochromatic CW signal regardless of its parametrization as long as we are 
able to describe the frequency evolution of the source in terms of a set of parameters $\lambda \in \mathbb{P}$.

\subsection{Distance implementation}

We proceed to explain how to adapt this newly introduced idea to the particularities of practical CW searches,
reserving a discussion on the actual effect of this distance as well as a comparison to previous approaches for 
sections \ref{sec:clustering} and \ref{sec:comparison}.

As already discussed, a first proposal of parameter-space distance can be constructed by comparing the
time-frequency tracks of a pair of candidates through the $L_{1}$ distance
\begin{equation}
    \label{eq:delta}
    d(\lambda, \lambda_{*}) \propto \int dt \abs{f_{\lambda}(t) - f_{\lambda_{*}}(t)}\;.
\end{equation}
While useful on its own, it can be improved by taking the following considerations into account.

First, the data products which all-sky CW searches deal with are discretized in a specific way.
For instance, most all-sky semicoherent searches work on Fourier-transformed data segments (so-called 
\emph{Short Fourier Transforms}, SFTs) with a time baseline $\Tsft$, which
in turn imposes a natural frequency discretization \mbox{$\delta f = \Tsft^{-1}$}. Each
$\Tsft$-long segment $\alpha$ is labeled by a starting \emph{timestamp} $t_{\alpha}$,
meaning we can get a discretized version of Eq.~\eqref{eq:delta} by introducing the substitution
\begin{equation}
    \label{eq:int_to_sum}
    \int dt \rightarrow \frac{1}{\delta f}\sum_{\alpha} \;,
\end{equation}
where the multiplying factors are kept in such a way that the distance stays 
dimensionless, i.e. the distance is measured in the corresponding (fractional) number of frequency bins $\delta f$.

Second, consider two templates $\lambda, \lambda_{*} \in \mathbb{P}$ such that their 
time-frequency tracks are, on average, $\tilde{\Delta}$ bins apart. If we assume a set of $N$ SFTs,
then from Eqs.~\eqref{eq:delta} and~\eqref{eq:int_to_sum} the distance is given by
$$
d(\lambda, \lambda_*) 
\propto \sum_{\alpha=0}^{N-1} \abs{f_{\lambda}(t_\alpha) - f_{\lambda_{*}}(t_\alpha)} 
\simeq \sum_{\alpha=0}^{N-1} \tilde{\Delta}
= N \tilde{\Delta}\;, $$
i.e. longer observing times would be able to build up higher distance values. We can factor this 
dependency out by choosing a proper normalization, effectively constructing a parameter-space distance
whose properties will be maintained across different observing runs:
\begin{equation}
    d(\lambda, \lambda_{*}) \equiv
    \frac{1}{N \delta f} \sum_{\alpha} \abs{f_{\lambda}(t_\alpha) - f_{\lambda_{*}}(t_\alpha)} \;.
    \label{eq:tbd}
\end{equation}
This last expression can be interpreted as the \emph{average mismatch} among the time-frequency tracks 
associated to a pair of templates.

Also, in the same way as search methods are able to produce sound results
for data streams containing gaps due to down time of the detectors, Eq.~\eqref{eq:tbd} can be computed using
a sub-set of timestamps $\bar{\alpha}$, reducing the amount of required computations
\begin{equation}
    d(\lambda, \lambda_{*}) \equiv \Tsft \cdot \langle \abs{f_{\lambda} - f_{\lambda_{*}}} \rangle\;,
    \label{eq:TBD}
\end{equation}
where $\langle \cdot \rangle$ represents an average over a certain set of timestamps
$\alpha \in \bar{\alpha}$. 

The following subsection introduces the main properties of Eq.~\eqref{eq:tbd} and compares them to those of
Eq.~\eqref{eq:TBD}. After showing their quantitative equivalence, we will simply refer to Eq.~\eqref{eq:TBD} 
as the \emph{Time-frequency Track Distance} (TTD).

\subsection{Discussion}

\begin{figure}
    \includegraphics[width=0.46\textwidth]{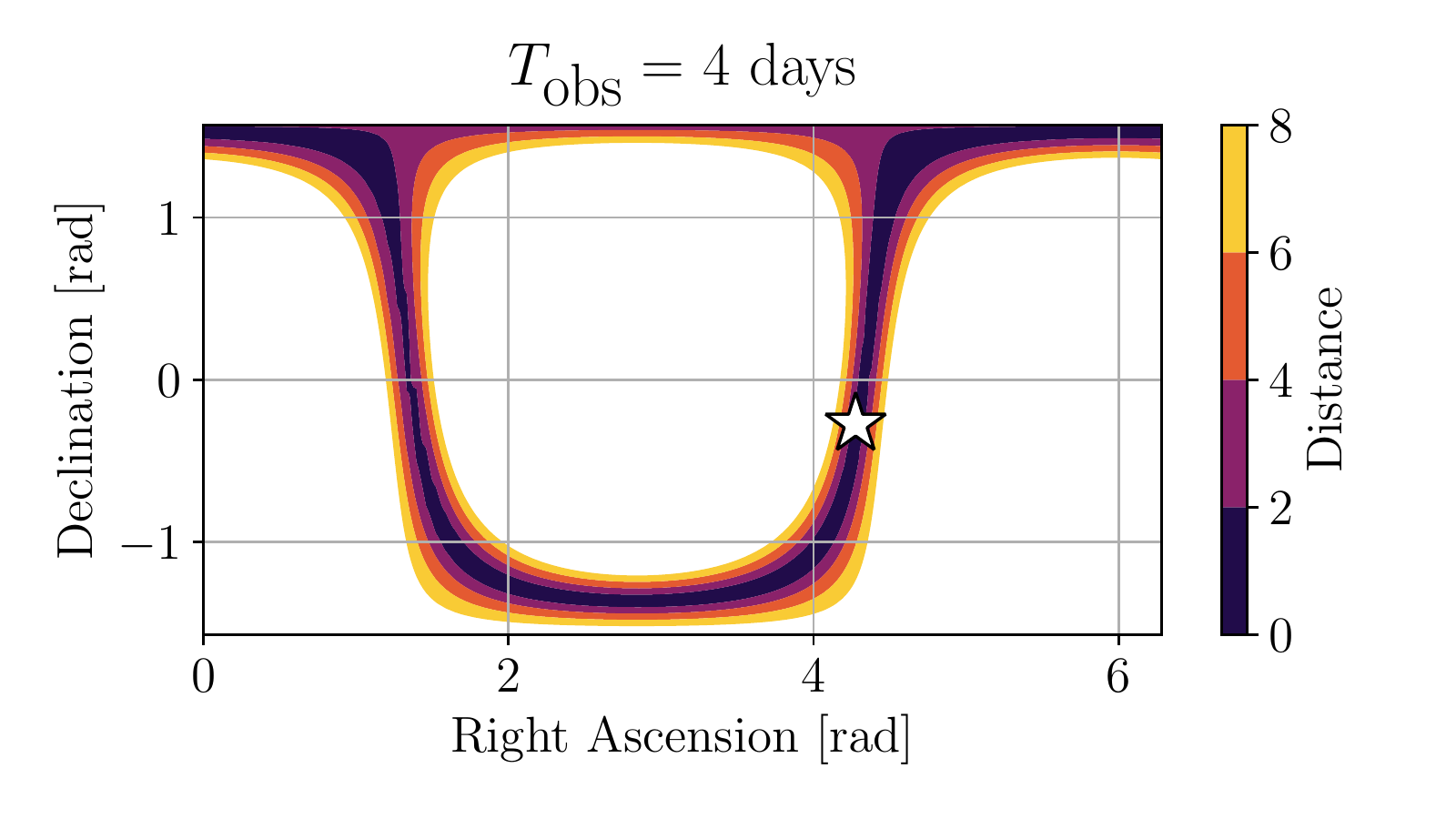}
    \includegraphics[width=0.46\textwidth]{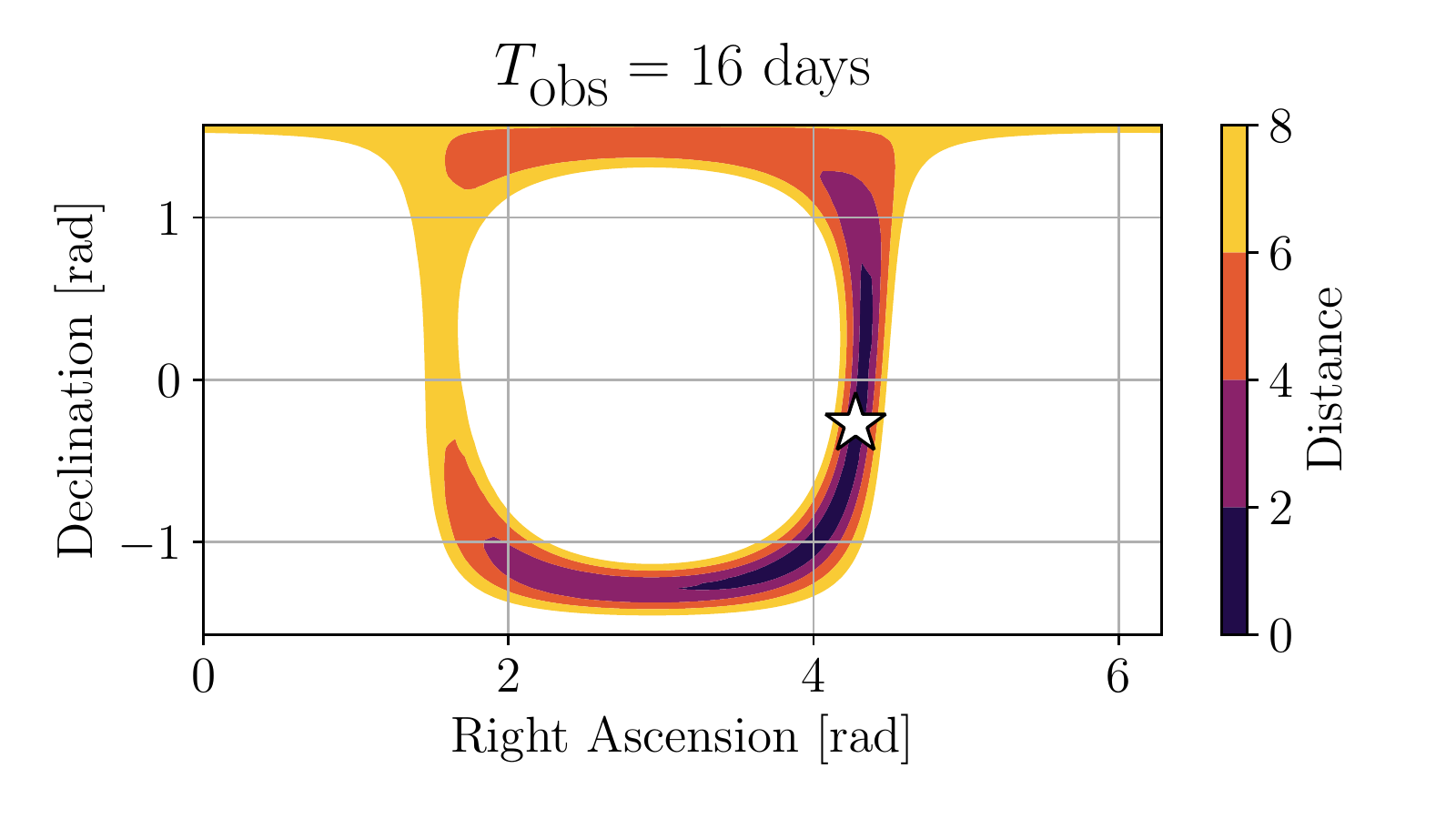}
    \includegraphics[width=0.46\textwidth]{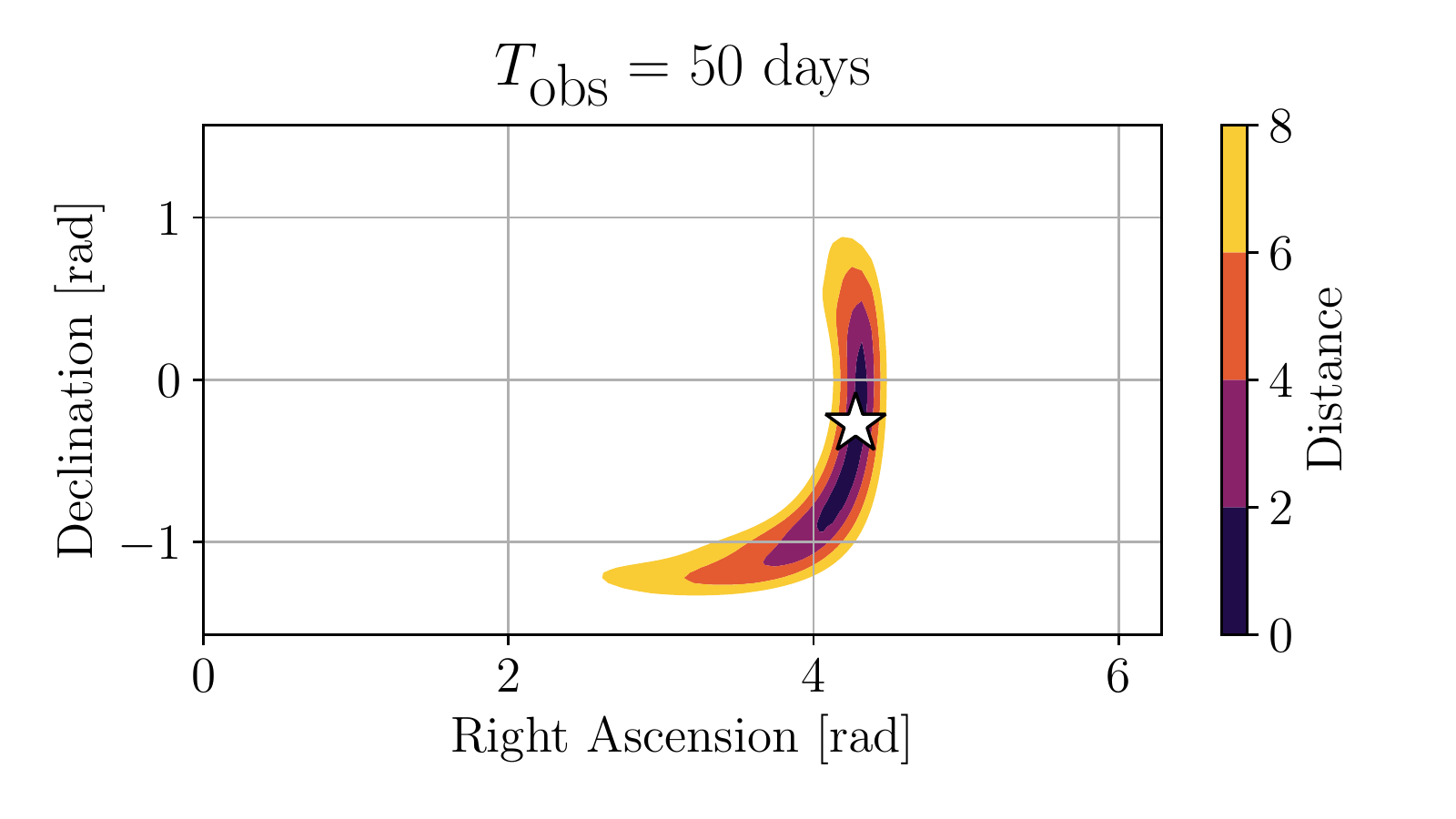}
    \includegraphics[width=0.46\textwidth]{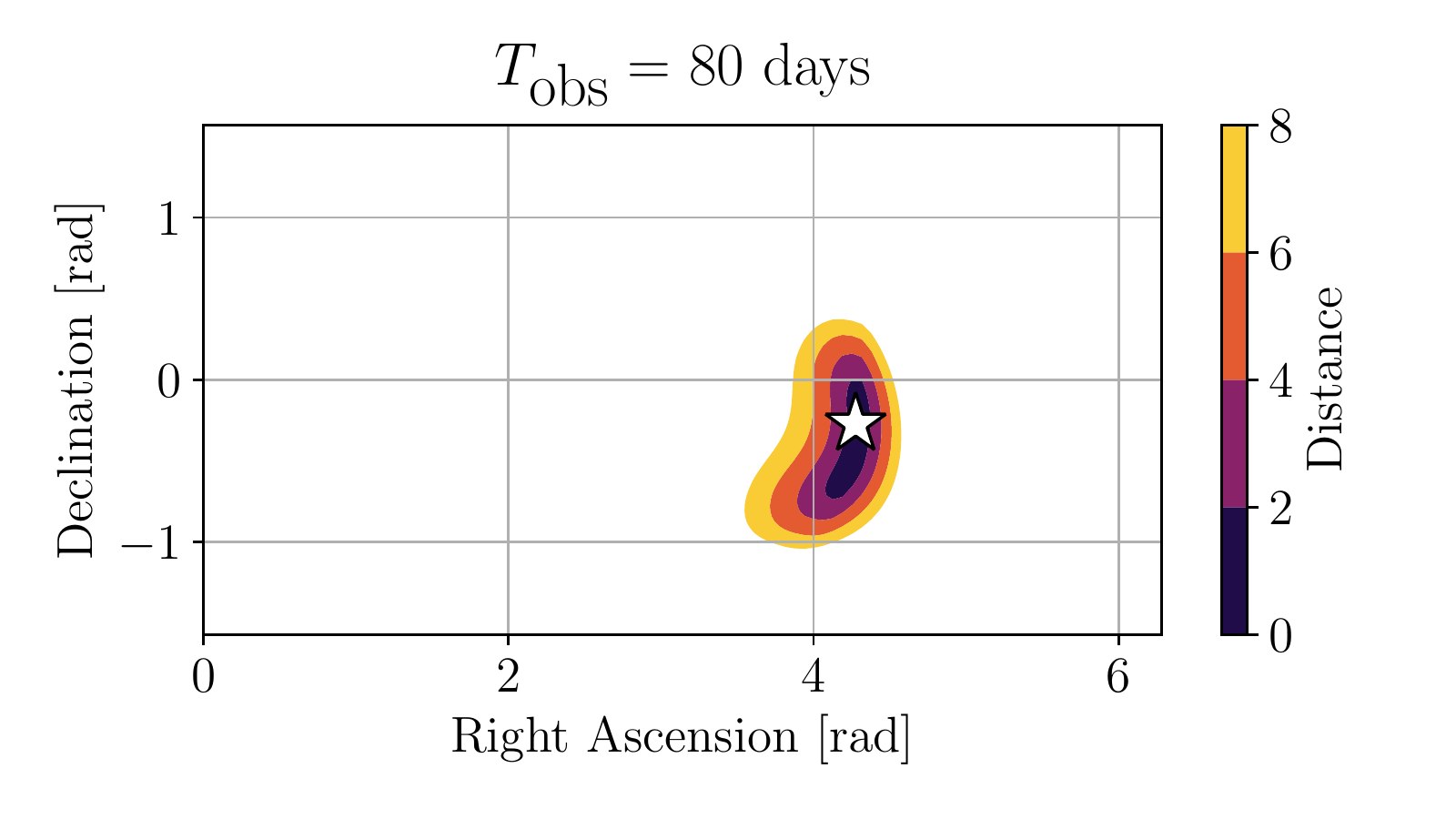}
    \caption{Evaluation of the distance from Eq.~\eqref{eq:tbd} across the sky with respect to 
    a reference sky position
    \mbox{$(\alpha, \delta) = (4.27, -0.27)\;\textrm{rad}$} for different observing times.}
    \label{fig:tobs_effect}
\end{figure}

To illustrate the properties of our new distance proposal from Eq.~\eqref{eq:tbd},
we first consider the simplest case of a CW signal, consisting of a source emitting GWs at a certain fixed frequency
$f_{0}$. Neglecting any proper motion of the source, the detected signal only experiences the Doppler modulation from the
detector's movement with respect to the solar system barycenter~\cite{PhysRevD.70.082001}:
\begin{equation}
    f_{\lambda}(t) = f_{0}\cdot \left(1 + \frac{\vec{v}(t)}{c} \cdot \vec{n}\right)\;,
\end{equation}
where $ v / c\simeq 10^{-4}$ is the detector velocity expressed in natural units and $\vec{n}$ represents the
sky position of the source. Focusing on sky position effects will deliver useful results for a general 
discussion of the implications of our distance, since this is the only common contribution to every kind of
CW search (as long as considering ground-based detectors).

We will make use of open data from the O2 observing run of the Advanced LIGO detectors \cite{O2Data, O2Datadoi}
to extract timestamps and detector velocity vectors using \cite{lalsuite}. 
Both LIGO detectors are taken into account by extending Eq.~\eqref{eq:tbd} to a multi-detector setup: 
track mismatches are computed for each detector,
computing an overall average over the timestamps of all used SFTs. This approach is possible due to the usage of
track mismatch rather than SNR loss to compare parameter-space points, as the former is
unaffected by amplitude modulations due to the different antenna pattern functions of each detector.

The parameter-space structure of this model is highly dependent on the length of the data stream under study.
As first pointed out in \cite{PhysRevD.70.082001}, if we measured the frequency of a CW signal
only at a certain moment in time, we would not be able to fully locate its corresponding sky position, since the
Earth-induced Doppler shift would point us to a \emph{circle} on the celestial sphere. It is through the
integration over multiple timestamps that the sky position of the source becomes well resolved.

This type of structure is well captured by Eq.~\eqref{eq:tbd}, as shown in Fig.~\ref{fig:tobs_effect}:
For short observing times, not being able to resolve sky position corresponds to the locus of close 
parameter-space points (as measured by the distance) being broadly extended along a circle on the celestial 
sphere. As observing time increases, this degeneracy starts to break down, with the set of close templates
finally narrowing down to a compact neighborhood around the actual sky position of the source.
This example demonstrates how our proposal is able to automatically capture an underlying parameter-space 
structure. We note that the improved sky localization with increasing observing times is
a generic feature of CW searches, as it is the result of identifying sky regions consistent with the long-term Doppler
modulation induced by the Earth's movement with respect to the SSB \cite{PhysRevD.70.082001}.
Longer observing times also increase search sensitivity by increasing the amount of SNR that can be accumulated by candidates.

\begin{figure}
    \includegraphics[width=0.5\textwidth]{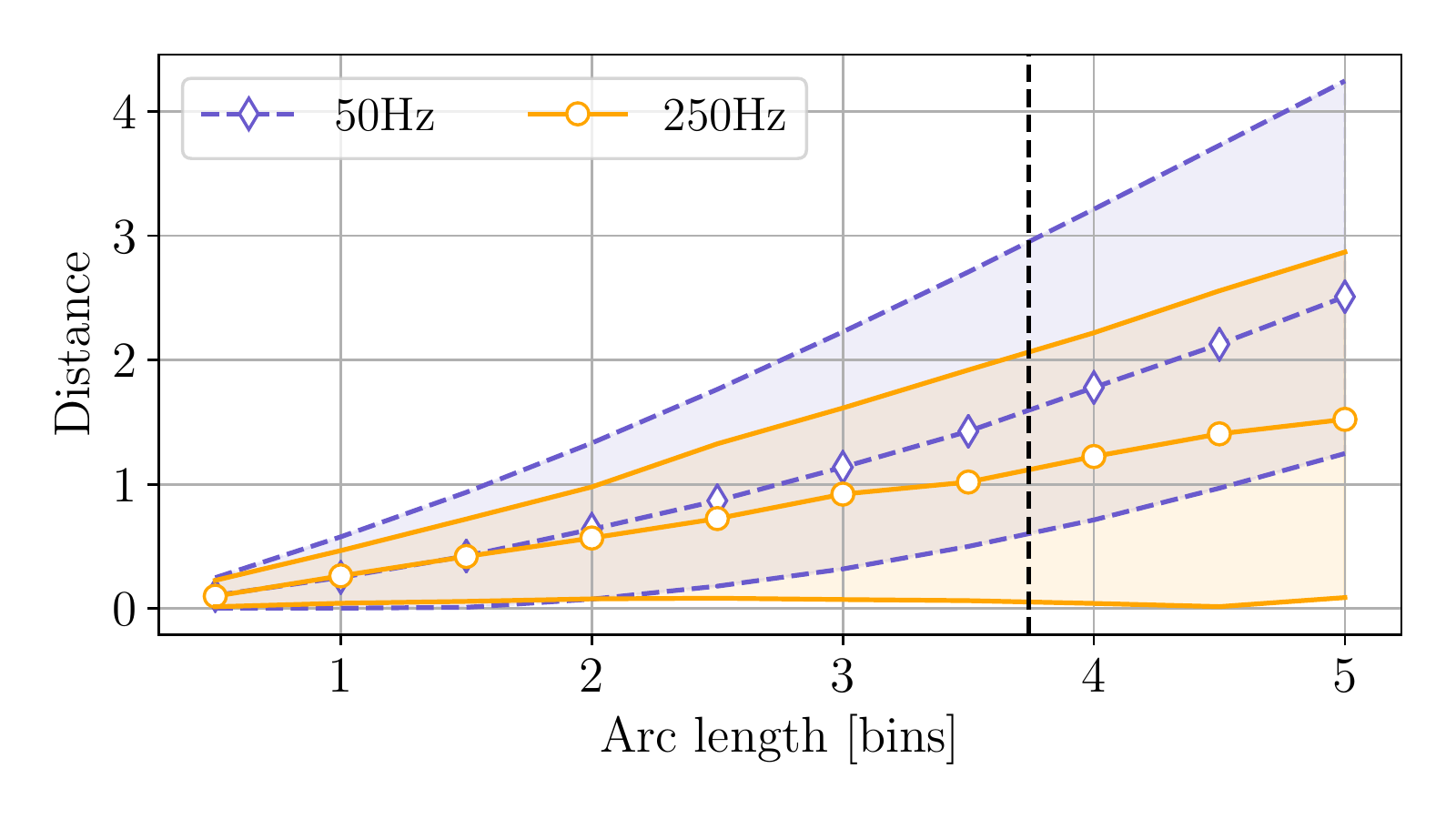}
    \caption{Distance comparison with respect to the great circle distance using $10^5$
    neighbouring sky positions, at two representative signal frequencies.
    Markers represent the average distance value for sky positions located at the
    specified number of bins $\pm 0.25$, and envelopes represent the maximum and minimum values within 
    each bracket. The vertical dashed line marks a value of $\sqrt{14}$ bins, the reference value employed by 
    \cite{PhysRevLett.124.191102} to cluster parameter-space points.}
    \label{fig:tbd_versus_bins}
\end{figure}

The behavior of Eq.~\eqref{eq:tbd} should be compared to those used in previous searches,
as discussed in section \ref{sec:current}.
In particular, we compare to the great circle distance to provide us with an estimation of the typical numerical scale of the new distance.
We express the great circle distance according to the natural sky resolution induced by the Doppler 
modulation (so called \emph{sky bins}) \cite{PhysRevD.70.082001}:
\begin{equation}
    \delta \theta = \left[\frac{v}{c} \cdot \Tsft \cdot  f_{0}\right]^{-1}\;.
    \label{eq:doppler_resolution}
\end{equation}
Both distances are compared in Fig.~\ref{fig:tbd_versus_bins}, using the same reference sky position as in
Fig.~\ref{fig:tobs_effect} (the actual choice of sky position plays a minor role, as we are focusing on
close neighborhoods in terms of the great circle distance). Given a certain distance value, higher frequencies
tend to correspond to longer circle lengths, as the bin resolution from Eq.~\eqref{eq:doppler_resolution} becomes thinner.
\begin{figure}
    \includegraphics[width=0.5\textwidth]{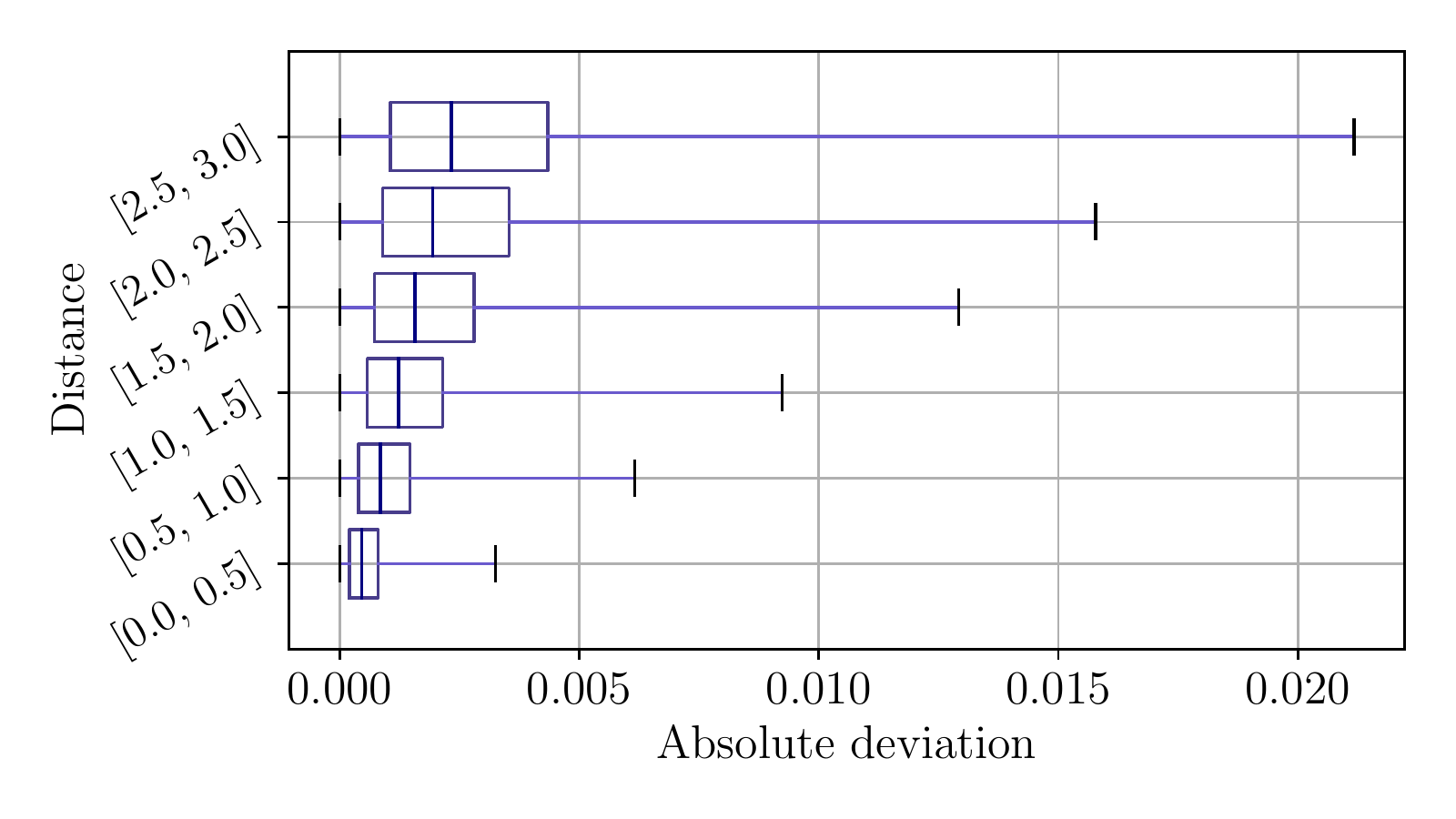}
    \caption{Distance comparison between Eqs.~\eqref{eq:tbd} and~\eqref{eq:TBD} using $3\times10^5$ pairs
    of parameter-space points corresponding to the binary parameter space of \cite{PhysRevLett.124.191102}.
    Vertical lines show the median of the distribution of values bracketed by the specified interval; 
    boxes mark the first and third quartiles of the distribution; whiskers denote the minimum and maximum 
    values within each of the brackets.} 
    \label{fig:subsampling}
\end{figure}

We note a distance of $\mathcal{O}(1)$ is enough to cover arc lengths corresponding to less than 4 parameter-space bins.
This will allow us to compare the performance of this distance to that of the Euclidean ansatz in
the context of a clustering algorithm in section \ref{sec:comparison}.

Next we test the validity of the reduced Eq.~\eqref{eq:TBD} with respect to Eq.~\eqref{eq:tbd} by computing the absolute
deviation produced by computing the distance using a subset of the available timestamps. Concretely, we select
500 timestamps along the O2 observing run data out of the initial $\mathcal{O}(10^{4})$.
In Fig.~\ref{fig:subsampling} we show the distribution of absolute deviations between Eq.~\eqref{eq:TBD} and Eq.~\eqref{eq:tbd},
using CW signals with binary modulations as in \cite{PhysRevLett.124.191102}.
Absolute deviations are two orders of magnitude smaller than the exact distance value within the useful regime;
hence we can safely use this approach to save on computing cost. This is consistent with the expected behavior of
time-frequency tracks: sufficiently close parameter-space points would produce similar time-frequency evolutions,
meaning their mismatch during a particular time span would be similar to the average value across the
whole run; hence, only a small fraction of the timestamps are required to be taken into account. This discussion 
is easily extendible to different types of sources, and can be taken at face value if relative motion effects are 
smaller than Earth-induced Doppler effects (which correspond to a relative frequency variation of $10^{-4}$).

\section{Application to clustering for an all-sky binary CW search}
\label{sec:clustering}

We now apply TTD to improve the clustering of candidates from CW searches.
This first example application will prove the benefits of using track-based parameter-space distances in CW searches.
Other possible uses of this proposal, such as the setup of parameter-space grids for the initial search stage,
or for the follow-up of candidates, are left for future work.

For the rest of this work, we will focus on the semicoherent \bsh pipeline~\cite{PhysRevD.100.024004, PhysRevLett.124.191102},
which is able to perform all-sky searches for CWs from unknown neutron stars in binary systems.
The pipeline selects a narrow frequency band (typically on the order of $0.1~\textrm{Hz}$)
and a certain binary parameter-space region, performing an all-sky
search and returning the most significant candidates according to a set of statistics.

We use the new distance in a clustering algorithm that is functionally equivalent to the one in \cite{PhysRevD.100.024004, PhysRevLett.124.191102},
which originally used the distance from Eq.~\eqref{eq:ecliptic};
from now on we refer to that reference algorithm as ``the O2 clustering''.
We comment on minor implementation differences in subsection~\ref{subsec:overview}, ensuring the fairness of the
comparison; a numerical assessment is postponed to Sec.~\ref{sec:comparison}.

The general starting point for clustering is a toplist $\mathcal{T}$ containing the most significant candidates of a certain
parameter-space region according to a search algorithm. We define a candidate as the pair $c^{(i)}=(\lambda^{(i)},
\mathfrak{S}^{(i)})$, where $\lambda^{(i)} \in \mathbb{P}$ represents the parameters associated to said candidate
and $\mathfrak{S}^{(i)} \in \mathbb{R}$ is the significance, represented by some statistic.

\subsection{Algorithm overview}
\label{subsec:overview}

The clustering algorithm consists of these steps:
\begin{enumerate}
    \item \textit{Construct reachable sets}: select candidates located in neighboring
        frequency bins to avoid having to compute irrelevant comparisons later on.\label{item:df}
    \item \textit{Construct coincidental sets}: compute TTD from Eq.~\eqref{eq:TBD} among the selected
        set of candidates and pair them according to a coincidence threshold 
        $\coincth$.\label{item:tbd}\footnote{Unlike in many other contexts of gravitational-wave data 
        analysis, by ``coincidence''
        here we are not referring to a comparison between data from different detectors.
        Instead, the point is whether two candidates coincide in their time-frequency evolution
        sufficiently closely.}
    \item \textit{Construct clusters}: group coincidental pairs into connected components. 
        Each connected component represents a \emph{cluster}.
    \item  \textit{Identify cluster centers}: retrieve the loudest candidate
        from each cluster according to a ranking statistic $\mathfrak{S}$.\label{item:center}
    \item \textit{Select clusters to follow up}: rank the clusters according to the statistic of
        their centers, selecting a certain number $N_{c}$ to follow up using a more sensitive 
        method.\label{item:rank}
\end{enumerate}
The first conceptual difference with respect to the O2 clustering is in step \ref{item:df}:
While we focus on the neighboring frequency bins of a candidate, the O2 clustering chains together neighboring
candidates until a frequency gap wider than a certain threshold is encountered. The main effect of such a 
strategy is to create wider reachable sets than strictly required, meaning the expected results are consistent 
with our proposal.

The second difference is related to the ranking criterion in step \ref{item:rank}, as the O2 clustering ranks
clusters according to the \emph{sum} of the individual statistics of their elements. In the presence of strong 
instrumental disturbances, individual candidates in the cluster are strong enough to be selected regardless of 
the exact criterion; in the case of signals (expected to be weak), however, there is an explicit dependence on the
number of toplist elements used, as selecting more candidates could lead to an inflation of the cluster population
due to the presence of candidates produced by pure noise. Selecting the \emph{loudest} candidate is safer in this 
regard, as the number of toplist elements plays no role. As discussed in \cite{BSH}, the effect of this 
difference will be negligible in our comparison as we will be focusing on estimating a sensitivity corresponding 
to the $95\%$ detection efficiency level.

Finally, the O2 clustering adds a minimum population to the selection criteria in step \ref{item:rank}: clusters must contain
at least 3 elements, being discarded otherwise. We comment on the effects of this criterion in Sec.~\ref{sec:comparison}.

The following subsections cover implementation details of each step of the algorithm. 

\subsection{Reachable sets}

We start by defining an auxiliary distance based only on the frequency difference (measured in bins)
between two candidates
\begin{equation}
    d_{f_0}(c, c_*) = \Tsft \cdot \abs{f_0^{c}-f_0^{c_*} }\;,
    \label{eq:aux}
\end{equation}
where $f_0^c$ represents the frequency associated to candidate $c$.
For each candidate $c \in \mathcal{T}$ we define its \emph{reachable set} as those candidates closer than 
a certain threshold $d^{\textrm{th}}_{f_0}$ according to Eq.~\eqref{eq:aux}:
\begin{equation}
    R[c] = \left\{ c_{*} \in \mathcal{T} \;|\; f_{0}^c \leq f_{0}^{c_*} \;\textrm{and}\;
    d_{f_0}(c, c_{*}) \leq d^{\textrm{th}}_{f_0} \right\}\;.
\end{equation}

This construction is motivated by the expected way a CW signal creates candidates 
across the parameter space, as the usual conventions taken by CW searches (and by \bsh~\cite{BSH} in particular)
aim to reduce its wandering to less than one frequency bin over a coherent segment;
hence, we can focus on a neighborhood of each candidate rather than the whole toplist. As a result,
in order to group candidates it is enough to focus on reachable sets.

\subsection{Coincidental sets}

Pair-wise distances using TTD are computed within each reachable set. Then,
we impose a \emph{coincidence threshold} $\coincth$ to define the \emph{coincidental set} of a candidate 
$c\in \mathcal{T}$ as those candidates within the reachable set whose distance to the reference candidate 
is lower than the specified threshold:
\begin{equation}
    C[c] = \left\{ c_{*} \in R[c] \;|\; d(c, c_{*}) \leq \coincth \right\}\;.
\end{equation}
Clusters are then constructed by computing the greatest groups of candidates such that any given candidate
within a group is closer than a distance of $\coincth$ to at least one other candidate within the same group.
This concludes the clustering procedure.

Each cluster possesses two attributes, namely a \emph{center} and a value of \emph{significance}.
The center of a cluster is defined as the \emph{loudest} point belonging to it, i.e., the one with the highest
detection statistic. Consequently, the \emph{significance} of a cluster is defined as that of the loudest point.

The resulting clusters are then selected according to their significance, using the cluster center as the
representative point to be furthered followed up using a more sensitive algorithm \cite{Shaltev:2014toa,
Aasi:2015rar, Papa:2016cwb, 2018pas7.conf...37S, PhysRevD.100.024004, Steltner:2020hfd, Ashton:2018ure, Keitel:2021xeq}.

\section{Improving the sensitivity of all-sky searches} 
\label{sec:comparison}

We assess the sensitivity improvement for an all-sky search due to applying our new distance
in the previously described clustering algorithm, using
a set of Monte-Carlo injections. This software injection campaign was performed using data from the Advanced 
LIGO O2 run so as to be comparable to the results obtained in the O2 open data all-sky search for CWs from neutron
stars in binary systems \cite{PhysRevLett.124.191102}. Clustering algorithms are functionally equivalent in
both cases, meaning sensitivity improvements can be attributed to the change of parameter-space distance.
We also discuss the choice of clustering parameters in detail in order to understand these improvements.

\subsection{The O2 data software injection campaign}
\label{subsec:O2data}
We reproduced the setup employed in \cite{PhysRevLett.124.191102} to obtain a faithful comparison to
their results. We used the full O2 Advanced LIGO run, spanning nine months of data taken by the two LIGO
detectors H1 (Hanford) and L1 (Louisiana) \cite{AdvancedLIGO, O2Data, O2Datadoi}. 
The employed time segments are those with the ``all'' tag in \cite{O2DataStamps}.
The data stream was divided into segments of duration \mbox{$\Tsft=900$\,s}
which were used to compute $50\%$ overlapping Fourier transforms (the previously introduced SFTs), as
discussed in \cite{PhysRevD.100.024004}; data from both detectors was analyzed together. This yields a set 
of 14788 SFTs from H1 and 14384 SFTs from L1.

Injections were performed at different \emph{sensitivity depth} values \cite{PhysRevD.91.064007, 
Dreissigacker:2018afk}
\begin{equation}
    \mathcal{D} = \frac{\sqrt{S_\mathrm{n}}}{h_{0}}\;,
\end{equation}
where $h_{0}$ represents the CW amplitude and $\sqrt{S_{n}}$ refers to an estimation of the amplitude 
spectral density of the detector noise; in essence, injecting at constant depth values allows us to produce
consistent injection sets with respect to the underlying noise floor, allowing for comparisons at different
frequency bands. 

We analyzed two different sets of injections. The first set aimed to understand the general behavior
of our proposal when used together with a clustering algorithm; the second one allowed us to assess the 
effectiveness of our proposal by obtaining an improvement in sensitivity. 

The generating procedure was the same for both sets. We first selected a set of representative frequency bands and
injected a certain number of CW signals into the data at different depths, drawing their parameters
from uniform distributions across the analyzed parameter space.
Then, we used the \bsh search to obtain a list of the most significant candidates to be clustered.
Following the criteria specified in \cite{PhysRevD.100.024004}, we counted an injection
as \emph{detected} if any of the \mbox{$N_{c} = 3$} most significant clusters containing at least three
candidates was significant enough so as to appear in the all-sky search toplist and each of its parameters were
no further than five parameter-space bins from the injection point.

\begin{table}
    \begin{tabular}{cc}
        \toprule
        Frequency Band [Hz] & $\langle \mathcal{D}^{95\%}\rangle \pm 1 \sigma$ \\
        \midrule
        $[100, 125)$ & $21.0 \pm 0.7$ \\
        $[125, 150)$ & $19.9 \pm 0.8$ \\
        $[150, 200)$ & $19.1 \pm 1.0$ \\
        $[200, 250)$ & $18.0 \pm 0.8$ \\
        $[250, 300)$ & $18.0 \pm 0.8$ \\
        \bottomrule
    \end{tabular}
    \caption{Summary of the average 95\% efficiency sensitivity depths obtained by the O2 open data all-
    sky search \cite{PhysRevD.100.024004}.}
    \label{table:depths}
\end{table}

Table \ref{table:depths} summarizes the sensitivity obtained by the O2 all-sky \bsh search
\cite{PhysRevLett.124.191102}. The quoted uncertainties correspond to the overall depth variance across the 
quoted frequency bands; these are consistent with (if not wider than) individual uncertainties obtained
through the fitting procedure, meaning the comparisons we are about to carry out do not overestimate the
effectiveness of the new distance.

For the first set of injections we selected two $0.1~\textrm{Hz}$ frequency bands,
namely $[150.1, 294.7]\,\textrm{Hz}$, and two sensitivity depth values, $[17.5, 21.5]\,\textrm{Hz}^{-1/2}$.
For each pair we injected 100 artificial CW signals.
Comparing with the $95\%$ efficiency obtained by \cite{PhysRevLett.124.191102}, a depth of
$21.5\;\textrm{Hz}^{-1/2}$ corresponds to a set of \emph{weak} injections, while $17.5\;\textrm{Hz}^{-1/2}$
corresponds to a set of \emph{strong} injections.
Lower frequencies tend to deliver greater $95\%$ efficiency depth values: the required parameter space resolution
becomes finer as we approach greater frequencies, as smaller Doppler modulations become resolvable by the method;
but in order to keep computing cost under control, coarser resolutions are used in practice, reducing the density
of the grid and, in turn, the effective parameter-space fraction under analysis.
This is a well-known fact affecting CW searches \cite{Wette:2013wza}.

For the second set we used six different frequency bands,
$[123.2, 129.8, 146.9, 165.2, 234.0, 262.7]\;\textrm{Hz}$, selecting six different depths
ranging from $14\;\textrm{Hz}^{-1/2}$ to $29\;\textrm{Hz}^{-1/2}$ (including a value of $32\;\textrm{Hz}^{-1/2}$ 
for frequencies below $200 \textrm{Hz}$). Here we injected 300 artificial signals at each sensitivity depth.

The clustering efficiency is given by the fraction of detected injections at each sensitivity 
depth. We will use the sensitivity depth corresponding to the $95\%$ efficiency, obtained through an interpolation
procedure, as a figure of merit for comparison purposes. A thorough exposition of this procedure can be found
in \cite{O1AllSky}.

\subsection{TTD: Clustering performance}
\label{sec:performance}
\begin{figure}
    \includegraphics[width=0.5\textwidth]{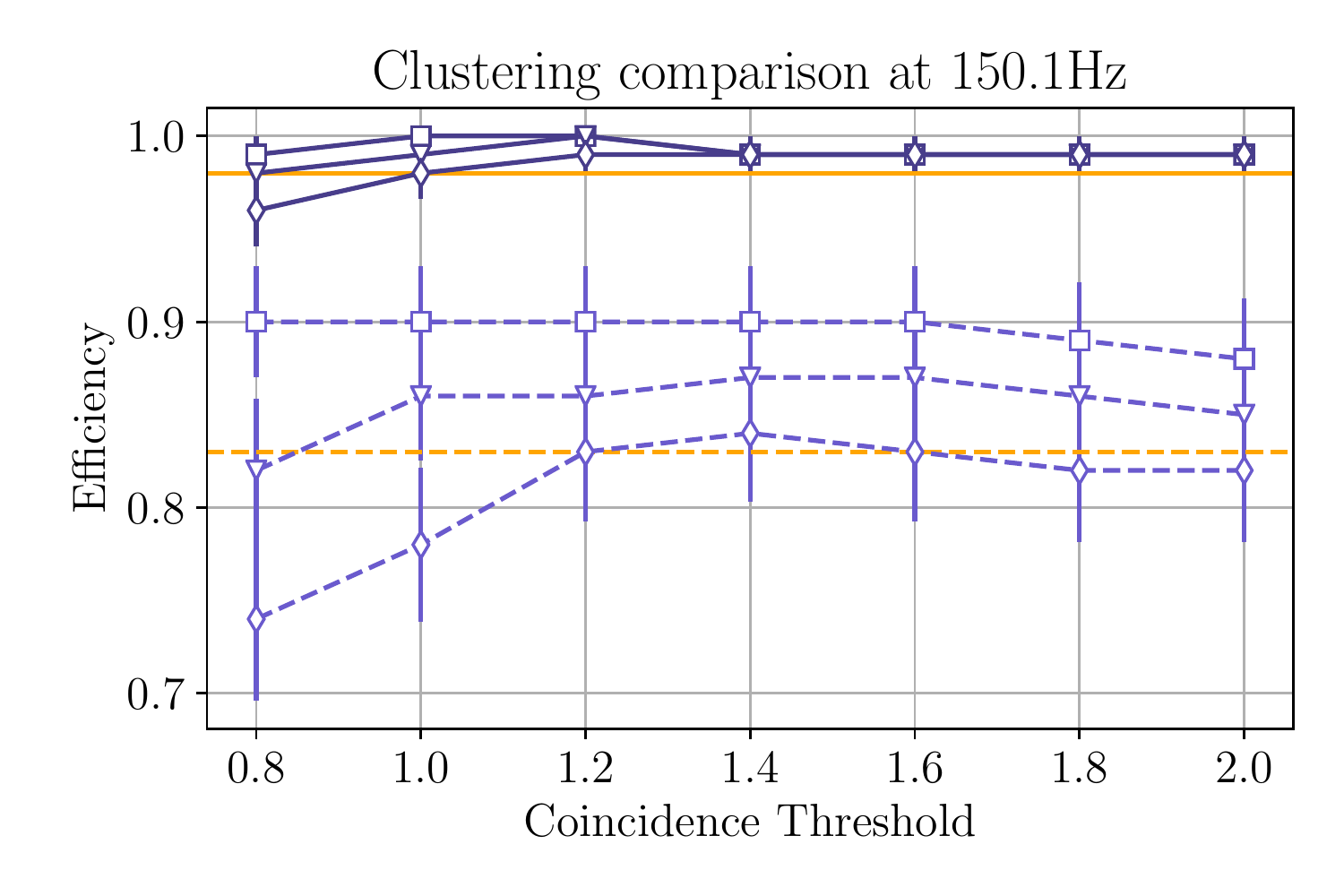}
    \includegraphics[width=0.5\textwidth]{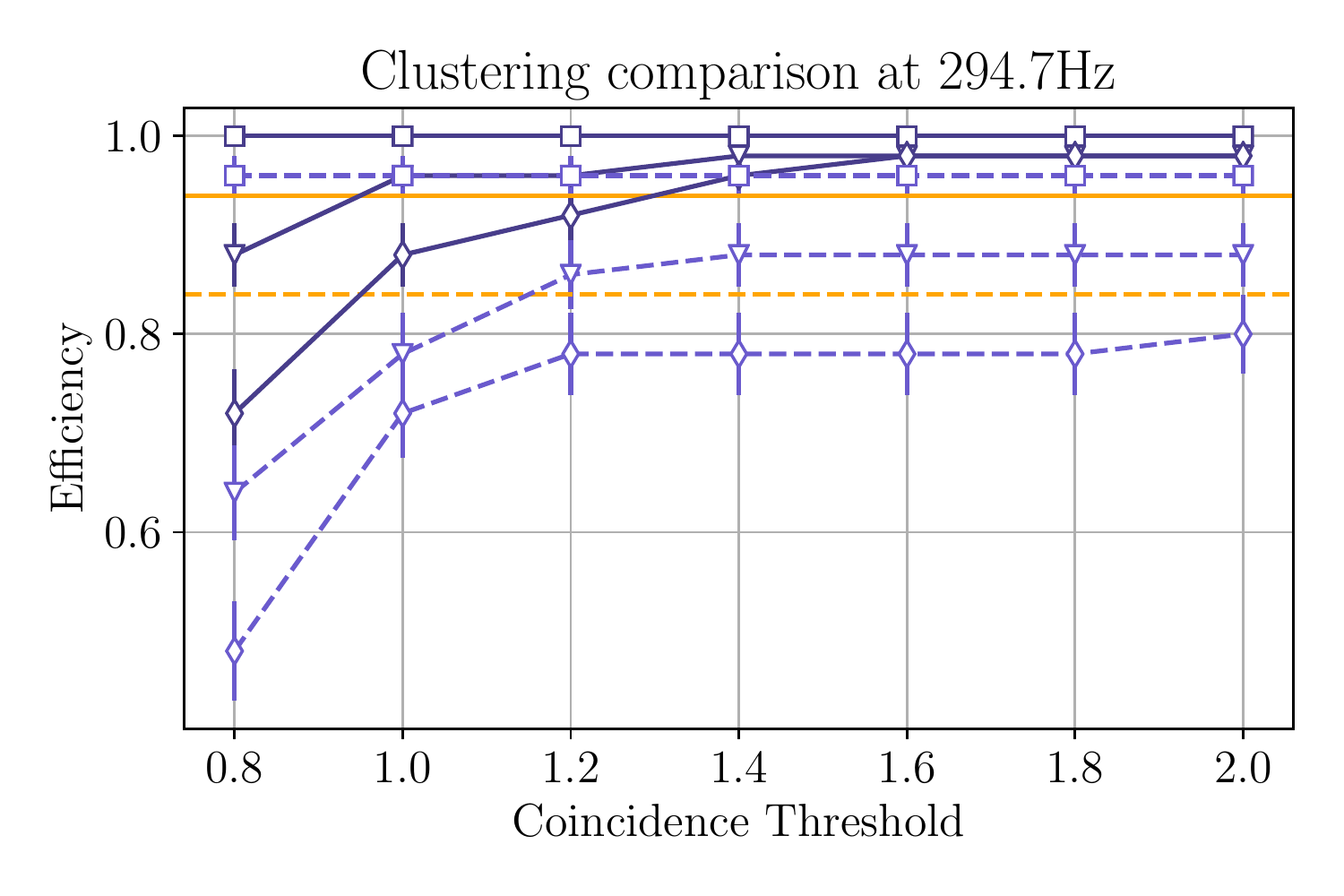}
    \caption{Performance comparison for different parameter choices, in two example frequency bands. 
    Solid lines represent the strong set of injections; dashed lines represent the weak set of injections. 
    Horizontal lines show the efficiency results obtained by the O2 clustering. 
    Markers represent the efficiency results obtained by using TTD.
    Square markers represent the results obtained by our proposal for different coincidence thresholds.
    Triangular markers show the results imposing a minimum cluster population of 3. 
    Diamond markers show the results imposing a minimum population of 5. 
    }
    \label{fig:efficiencies}
\end{figure}

First we discuss the clustering results obtained from the first set of injections.
The O2 clustering employed a distance threshold of \mbox{$d^{\textrm{th}}_{\textrm{O2}} = \sqrt{14} \simeq 3.7$} 
parameter-space bins, optimized through the use of Monte-Carlo injections \cite{PhysRevLett.124.191102};
efficiencies are reported using horizontal lines in Fig.~\ref{fig:efficiencies}.

A fair comparison requires us to understand how each of the clustering parameters, namely the thresholds to
construct reachable and coincidental sets and the minimum population threshold, interact with our proposal.

First, let us focus on the impact of changing the reachability threshold $\reachth$. This parameter controls the
number of neighboring frequency bins taken into account to compute distances between pairs of candidates:
as previously discussed, a CW candidate would tend to spread only across a small set of neighboring
frequency bins, either due to spectral leakage (if the CW parameters fell close to the border of a frequency bin) 
or parameter-space correlations; hence $\reachth$ could be kept at low values, incidentally reducing the
computing cost of the overall algorithm. Indeed, running the same setup for three reachability thresholds 
$\reachth = (1, 2, 3) \;[\Tsft^{-1}]$ showed no preferred value in terms of detection efficiency. As a result,
it is feasible to work with $\reachth = 1 \;[\Tsft^{-1}]$, reducing the computational cost by neglecting redundant
parameter-space relations.

Second, consider the interplay between a minimum population threshold (i.e. a minimum number of candidates
to define a cluster as such) and the coincidence threshold. In Fig.~\ref{fig:efficiencies} we show the efficiency
achieved by different combinations of these two parameters. Two main features are worth of comment:

First, the impact of a minimum population threshold is noticeable when combined with low coincidence thresholds:
Significant candidates are unable to group with their neighbors and, consequently, get removed due to the
lack of population in the cluster. As a result, potentially detectable signals (in the sense that the main stage
of the algorithm did \emph{not} reject them) are completely discarded. This effect is more noticeable for a set 
of weak injections, as they are less prone to enhance detection statistics across a wide region of parameter space;
higher frequencies are also affected by this more, since the density of parameter-space templates at those regions
tends to be lower than at low frequencies.

\begin{figure}
    \includegraphics[width=0.5\textwidth]{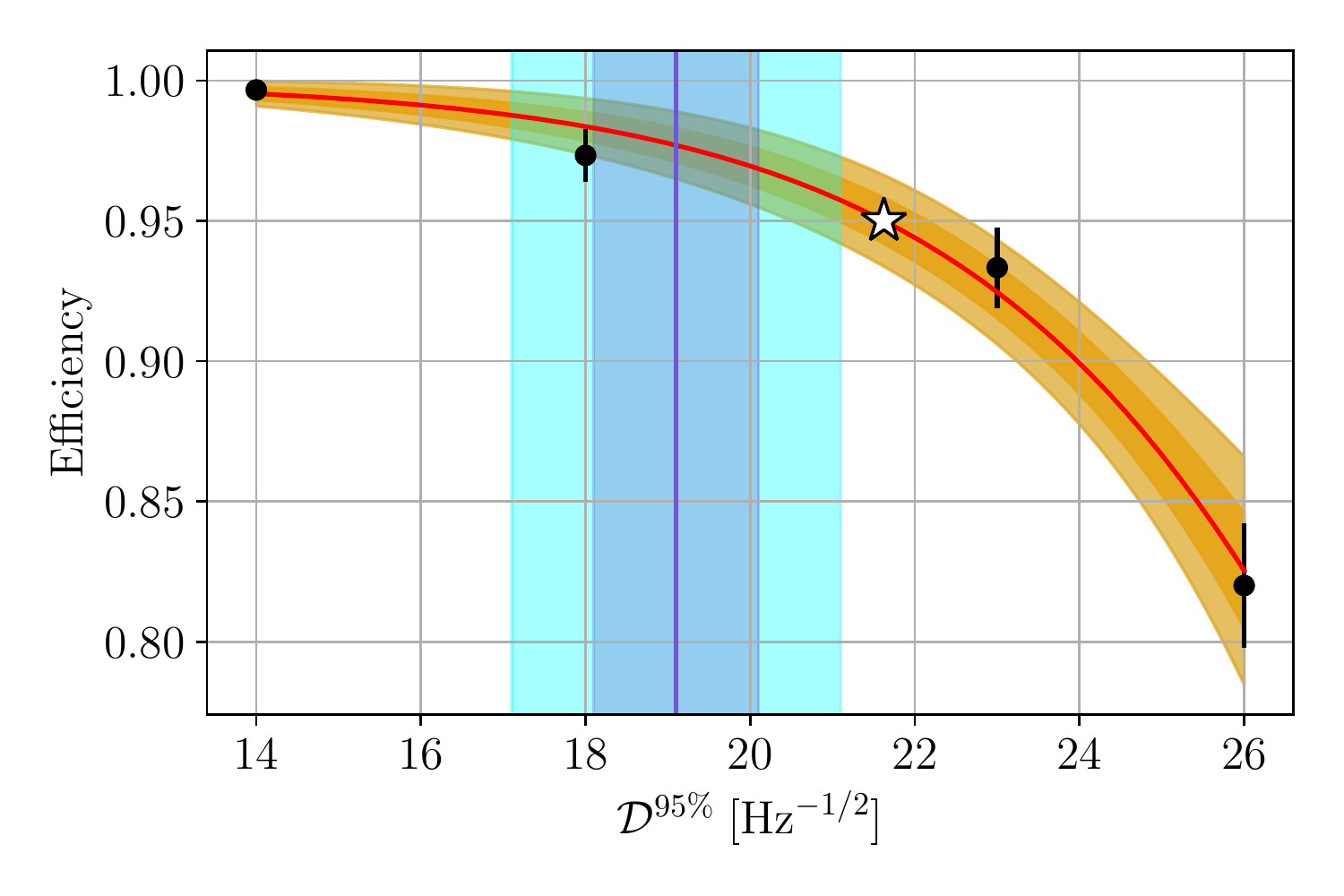}
    \caption{Example of 95\% efficiency depth interpolation at 165.2\,Hz. Points represent the detected
    fraction of 300 software injections after running a clustering using TTD with $\reachth = 1
    \;[\Tsft^{-1}]$ and $\coincth = 1.0$; the error bar on each point represents a binomial uncertainty.
    No minimum population is imposed. The vertical line shows the average sensitivity depth obtained by the O2 
    open data search at this frequency band. 
    Shaded regions represent $1\sigma$ and $2\sigma$ uncertainties.
    The 95\% efficiency depth interpolated from a sigmoid fit is represented by a star marker.}
    \label{fig:fit_example}
\end{figure}

Secondly, increasing the coincidence threshold partially mitigates this effect, as stand-alone significant 
parameter-space points are able to group together with their less significant neighbors.
Low-frequency results, however, suggest a limiting factor to this strategy, as this increase could also lead
to neighboring clusters merging together; given that the algorithm focuses on the \emph{center} of a cluster
only, this means a signal could be discarded if its cluster gets merged with an instrumental disturbance. 
This phenomenon is highly dependent on the data under analysis, as well as the properties of the CW signal 
in it. In particular highly disturbed bands tend to deliver highly significant clusters
corresponding to instrumental artifacts, as observed in previous all-sky searches \cite{O1AllSky, 
O1FullBandAllSky, PhysRevD.100.024004, 2018PhRvD..97h2002C}. Nevertheless, these clusters could also be 
isolated enough from the rest of interesting local maxima so as to avoid any mergers.

Let us impose a minimum population of three elements per cluster akin to the O2 clustering. Fig.~\ref{fig:efficiencies}
shows how TTD obtains improved results with respect to the Euclidean ansatz used in the original O2 algorithm, provided 
we select a suitable coincidence threshold. We devote the following subsection to expose the soundness of this improvement.

\subsection{TTD: Sensitivity improvement}

\begin{figure}
    \includegraphics[width=0.5\textwidth]{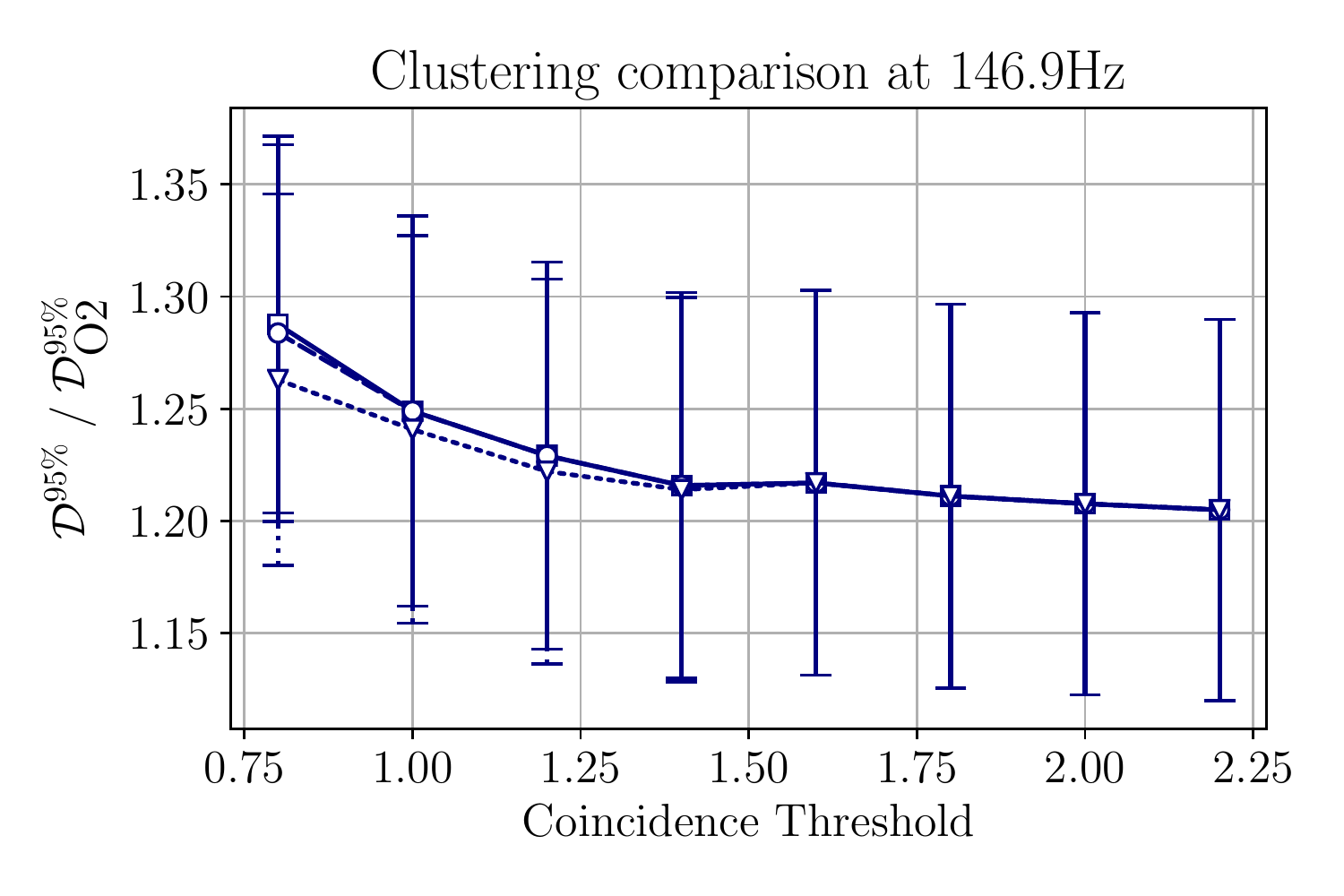}
    \includegraphics[width=0.5\textwidth]{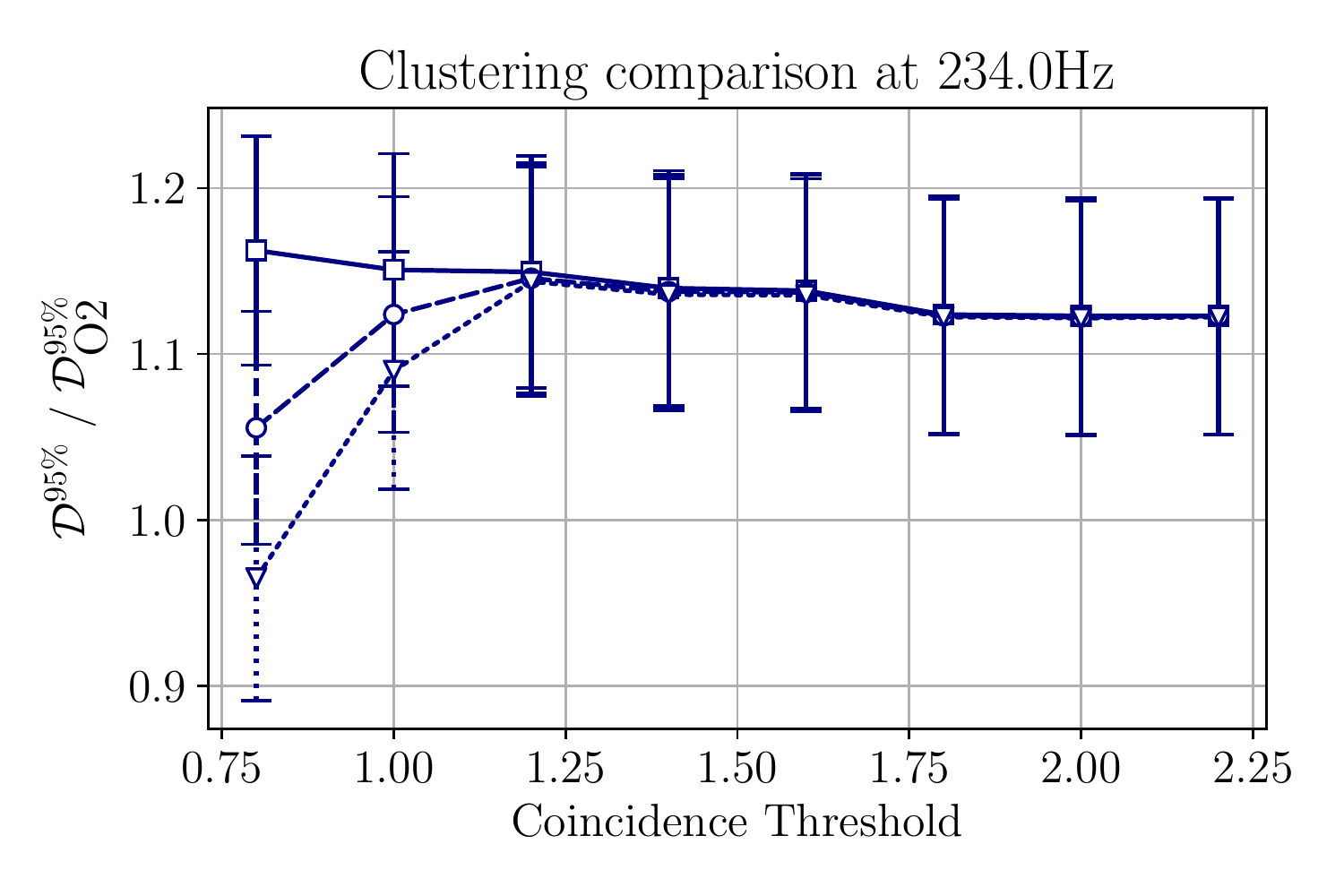}
    \caption{Relative sensitivity improvement at two particular frequency bands
    due to the usage of TTD as a function of the coincidence threshold and minimum 
    population parameters. Squares and a solid line represent no population threshold;
    circles and a dashed line represent a minimum population of one;
    inverted triangles and a dotted line represent a minimum population of three.
    \label{fig:depth_improvement_bands}
    }
\end{figure}

\begin{figure}
    \includegraphics[width=0.5\textwidth]{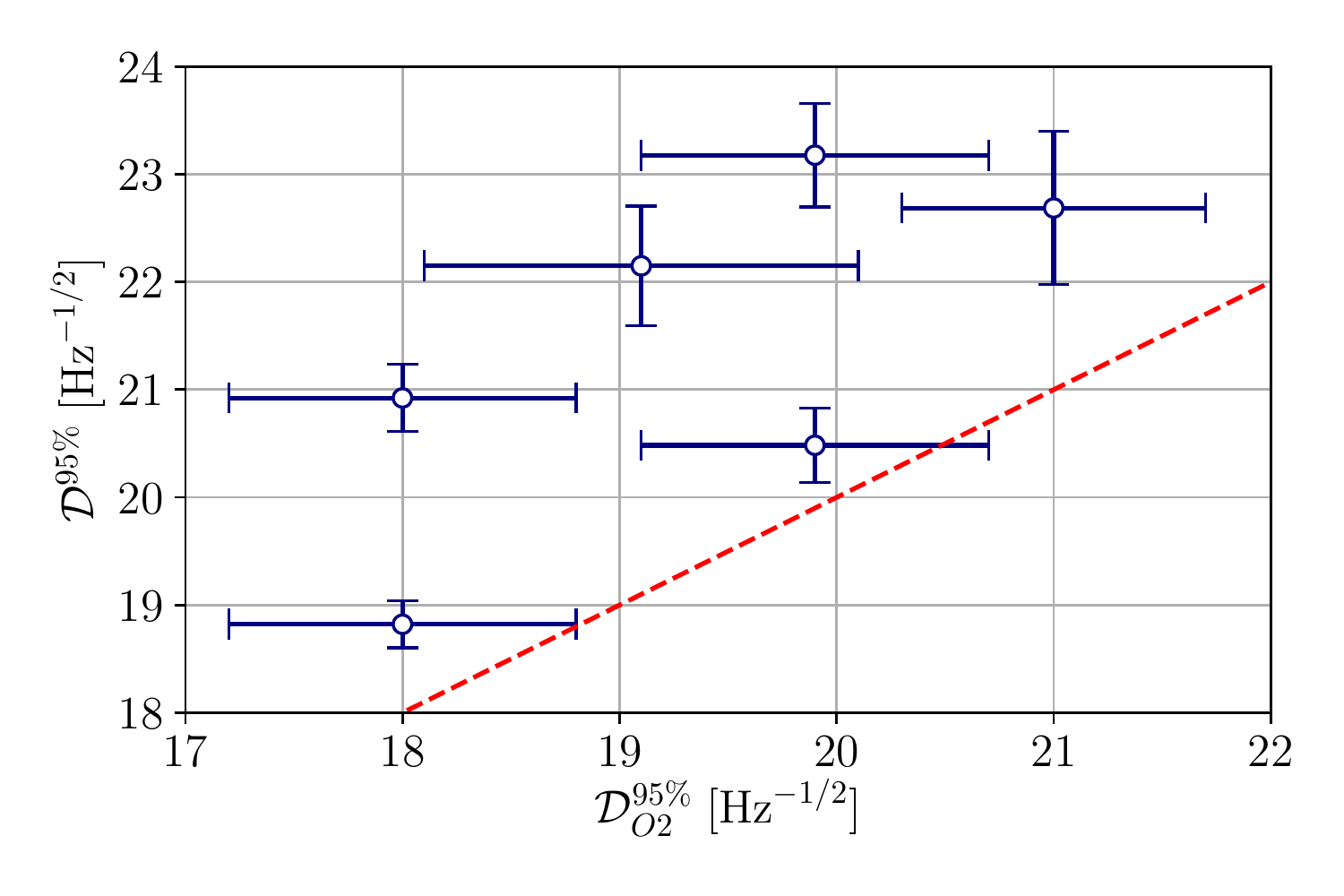}
    \caption{Comparison of 95\% sensitivity depths from the second set of CW software injections
    obtained using TTD against depths from~\cite{PhysRevLett.124.191102}.
    Each point corresponds to the depth estimated from injections in one frequency band.
    The dashed line represents equality. Vertical uncertainties
    are derived from the fitting procedure as illustrated in Fig.~\ref{fig:fit_example}.
    Horizontal uncertainties are those reported in Table~\ref{table:depths}.
    \label{fig:depth_improvement}
    }
\end{figure}
With the second set of injections we provide an assessment of the effectiveness of our proposal in
terms of search sensitivity. We recall this set consisted of five frequency bands, selecting six sensitivity 
depths for each of them, at which we injected 300 CW signals each. We use these injections to interpolate the
sensitivity depth corresponding to $95\%$ detection efficiency, $\mathcal{D}^{95\%}$. This is a commonly used
figure of merit in CW searches. A higher depth values means being able to detect weaker signals, as they are
buried deeper in the noise. This procedure is customarily employed in CW searches in order 
to assess their sensitivity \cite{Walsh:2016hyc, O1AllSky, O1FullBandAllSky, PhysRevD.100.024004, 
PhysRevLett.124.191102}. An example of the interpolation procedure for one band is provided in Fig.~\ref{fig:fit_example}.

Our previous discussion on the implications of a population threshold can be extended using the
95\% detection efficiency depth. In Fig.~\ref{fig:depth_improvement_bands} we compare the relative
improvement obtained with respect to the O2 clustering by simply using TTD from Eq.~\eqref{eq:TBD} rather than
Eqs.~\eqref{eq:euclidean}, \eqref{eq:ecliptic} and \eqref{eq:euclidean_binary}.

By properly tuning the coincidence threshold, the usage of our proposal under the same conditions as the O2 clustering
achieves a better sensitivity. As already discussed, the application of a population threshold can be compensated 
by a higher coincidence threshold. The specific increase notably depends on the
frequency band (and more concretely, on the template density of the parameter-space region) under study.

We summarize the improvement in sensitivity in Fig.~\ref{fig:depth_improvement},
comparing our 95\% efficiency sensitivity depth estimates against the results reported
in Table~\ref{table:depths}. Results using TTD are maximized with respect to the employed
coincidence threshold, deeming them to be insensitive to the imposition of a minimum population threshold.
We report a $5-15\%$ improvement by the sole use TTD of instead of an Euclidean ansatz.

\section{Conclusion}
\label{sec:conclusion}

We have introduced a new parameter-space distance for continuous gravitational-wave searches
by relating parameter-space points to their corresponding instantaneous frequency evolution.
As opposed to previous approaches, this proposal provides a simple way of comparing candidates
from CW semicoherent searches in a consistent fashion with respect to the underlying parameter-space
structure to which the searches themselves are sensitive.

We have demonstrated this consistency of the new distance with the well-known behavior of semicoherent
searches by reproducing a well-understood parameter-space structure related to the
Earth-induced Doppler modulation of ground-based detectors, the ``circles in the sky''.
The new distance can be computed as an average over a subset of timestamps from a given data set,
significantly reducing its computational cost at negligible numerical accuracy loss.

Our proposal is effective in improving practical CW searches, as we have demonstrated by using it to improve
the candidate clustering procedure from the O2 open data search for CWs in binary systems of~\cite{PhysRevLett.124.191102}
and reproducing their sensitivity estimation procedure. After illustrating
the role played by different parameters in the procedure, with the best settings we have obtained a $5-15\%$ sensitivity
improvement from using the new distance in the clustering step.

This distance definition can be seamlessly extended to any kind of quasi-monochromatic CW signal, provided we 
are able to describe its associated frequency evolution, as the main operational quantity is not provided by 
parameter-space points, but by their embedding into the time-frequency plane according to the CW model under
consideration. This makes it a valuable tool to develop new post-processing stages and improve the sensitivity of wide parameter-space searches.

\section*{Acknowledgements}
We gratefully thank P. B. Covas, Evan Goetz, Luca Rei and the LIGO-Virgo-KAGRA Continuous Wave working group
for many fruitful discussions. We would like to thank the anonymous referee for their useful suggestions to
improve the quality of this document.
This work was supported by European Union FEDER funds, the Ministry of Science, 
Innovation and Universities and the Spanish Agencia Estatal de Investigaci\'on grants
PID2019-106416GB-I00/AEI/10.13039/501100011033,  
FPA2016-76821-P,     
RED2018-102661-T,    
RED2018-102573-E,    
FPA2017-90687-REDC,  
Comunitat Autonoma de les Illes Balears through the Direcci\'o General de Pol\'itica Universitaria i Recerca with funds from the Tourist Stay Tax Law ITS 2017-006 (PRD2018/24),
Generalitat Valenciana (PROMETEO/2019/071),  
EU COST Actions CA18108, CA17137, CA16214, and CA16104.
R.~T.~is supported by the Spanish Ministerio de Ciencia, Innovaci{\'o}n y Universidades (ref.~FPU 18/00694).
D.~K.~is supported by the Spanish Ministerio de Ciencia, Innovaci{\'o}n y Universidades (ref.~BEAGAL 18/00148)
and cofinanced by the Universitat de les Illes Balears.
The authors thankfully acknowledge the computer resources at CTE-Power and the technical support provided by 
Barcelona Supercomputing Center - Centro Nacional de Supercomputaci\'on through grant 
No. AECT-2019-3-0011 from the Red Espa\~nola de Supercomputaci\'on (RES).

This research has made use of data, software and/or web tools obtained from the Gravitational Wave Open Science Center (https://www.gw-openscience.org/), a service of LIGO Laboratory, the LIGO Scientific Collaboration and the Virgo Collaboration. LIGO Laboratory and Advanced LIGO are funded by the United States National Science Foundation (NSF) as well as the Science and Technology Facilities Council (STFC) of the United Kingdom, the Max-Planck-Society (MPS), and the State of Niedersachsen/Germany for support of the construction of Advanced LIGO and construction and operation of the GEO600 detector. Additional support for Advanced LIGO was provided by the Australian Research Council. Virgo is funded, through the European Gravitational Observatory (EGO), by the French Centre National de Recherche Scientifique (CNRS), the Italian Istituto Nazionale della Fisica Nucleare (INFN) and the Dutch Nikhef, with contributions by institutions from Belgium, Germany, Greece, Hungary, Ireland, Japan, Monaco, Poland, Portugal, Spain.

 This paper has been assigned document number LIGO-P2000493.

\bibliography{references} \bibliographystyle{apsrev4-2}

\end{document}